\documentclass[showpacs,preprintnumbers,amsmath,amssymb, nofootinbib]{revtex4}

\usepackage{pictex, dcpic}
\usepackage{graphicx}
\usepackage{dcolumn}
\usepackage{bm}

\def\al{\alpha}
\def\be{\beta}
\def\de{\delta}

\def\ep{\epsilon}

\def\te{\theta}
\def\la{\lambda}

\def\om{\omega}
\def\si{\sigma}
\def\vp{\varphi}

\def\ka{\kappa}

\def\De{\Delta}
\def\Ga{\Gamma}

\def\Om{\Omega}

 \def\calR{{\hbox{\cal R}}}

\def\gotC{{{\mathfrak C}}}
\def\gotP{{{\mathfrak P}}}
 \def\gotP{{{\mathfrak P}}}
 
\def\id{{\hbox{\rm id}}}

\def\ip{\hbox to4pt{\leaders\hrule height0.3pt\hfill}\vbox to8pt{\leaders\vrule width0.3pt\vfill}\kern 2pt}
 
\def\del{\partial}
\def\na{\nabla}

\def\Lie{\hbox{\LieFont \$}}

\def\arr{\rightarrow}

\def\then{\Rightarrow}

\def\nab#1{{\buildrel #1\over \na}}
\def\lsim{\hbox{\lower.7ex\hbox{${\buildrel< \over \sim}$}}} 

\def\Frac[#1/#2]{\frac{#1}{#2}}
\def\({\left(}
\def\){\right)}
\def\[{\left[}
\def\]{\right]}
\def\^#1{{}^{#1}_{\>\cdot}}
\def\_#1{{}_{#1}^{\>\cdot}}
\def\Label=#1{{\buildrel {\hbox{\fiveSerif \ShowLabel{#1}}}\over =}}
\def\<{\kern -1pt}

\def\Dal{\hbox{\tenRelazioni  \char003}}

\def\[{\begin{equation}}
\def\]{\end{equation}}
\begin{document}
\def\Lie{\pounds}
\def\Dal{\Box}

\newpage

\title{Mathematical Equivalence vs. Physical Equivalence \\ between Extended Theories of Gravitations}

\author{Lorenzo Fatibene}
 \email{lorenzo.fatibene@unito.it}
  \affiliation{Dipartimento di Matematica, Universit\`a di Torino, Italy\\
 INFN Sezione Torino- Iniz.~Spec.~Na12}

\author{Mauro Francaviglia}
 \email{mauro.francaviglia@unito.it}
  \affiliation{Dipartimento di Matematica, Universit\`a di Torino, Italy\\
 INFN Sezione Torino- Iniz.~Spec.~Na12}


\begin{abstract}
We shall show that although Palatini $f(\calR)$-theories are equivalent to Brans-Dicke theories, 
still the first pass the Mercury precession of perihelia test, while the second do not.
We argue that the two models are not physically equivalent due to a different assumptions about  free fall.

We shall also go through perihelia test without fixing a conformal gauge (clocks or rulers) in order to highlight what can be measured in a conformal invariant way and what cannot.    
We shall argue that the conformal gauge is broken by choosing a definition of clock, rulers or, equivalently, of masses. 
\end{abstract}


\maketitle

\def\FieldEqsEquivalenceAPP{A}
\def\FrameAPP{B}
\def \ProjectionIdentitiesAPP{C}

\section{Introduction}

In the early 70s Ehlers-Pirani-Schild  (EPS) proposed an axiomatic approach to gravitational physics; see \cite{EPS}.
They decided to start from potentially observable quantities, namely the families of worldlines of massive particles and light rays,
 and define out of them the geometry of spacetime.
They assumed properties of these families together with their mutual relation, that are physically reasonable and well motivated in the classical regime.

The final output is that geometry of spacetime is described by a {\it conformal structure},
i.e.~a class $\gotC=[g]$ of Lorentzian metrics
\[
\gotC=[g]=\{\tilde g= \Phi^2 \cdot g\}
\] 
together with a {\it projective structure},
i.e.~a class $\gotP=[\Ga]$ of connections
\[
\gotP=[\Ga]=\{\tilde\Ga^\al_{\be\mu}= \Ga^\al_{\be\mu}+ A_{(\mu}\de^\al_{\be)} :\> A_\mu\>\hbox{a 1-form}\}
\] 

The conformal structure can be used to define {\it timelike, lightlike, spacelike} directions in spacetime as well as  light cones.
Let us stress, however, that $\gotC$ does not define a notion of length along spacelike or timelike curves. Length depends on a representative of the conformal class
$\gotC$, i.e.~on a specific Lorentzian metric $g$ in $\gotC$. The other way around, if one has $\gotC$ and defines a notion of length, then this notion singles out precisely a representative in the given conformal structure.

The different connections in a projective structure $\gotP$ do define the same geodesics (or, better, {\it autoparallel}) trajectories, though differently parametrized, as well as the same free falling of massive particles.
(Let us remark that a trajectory in spacetime parametrized in two different ways does in fact represent essentially the same physical motion; parametrization of timelike worldlines is related to clocks and it is in fact a {\it convention} unless absolute time exists.)

Since light is deflected by Gravity as it happens for massive particles one expects the conformal and projective structures to be somehow related.
The conformal and projective structures are said to be {\it EPS-compatible} when lightlike geodesics of the conformal structure form a subset of geodesic trajectories of the projective structure. 
Given a pair $(\gotC, \gotP)$ of compatible conformal and projective structures, 
one can canonically fix a representative $\tilde \Ga$ of the projective structure $\gotP$ such that the following holds
\[
\nab{\tilde\Ga}_\mu g_{\al\be}= 2 A_\mu g_{\al\be}
\label{EPSC}\]
for some $1$-form $A_\mu$.
If condition (\ref{EPSC}) holds true (i.e.~it exists $A$ such that (\ref{EPSC}) holds true for the given $g\in \gotC$) then for any other representative $\tilde g\in \gotC$ there exist a $1$-form $\tilde A$ as well such that (\ref{EPSC}) singles out the same connection $\tilde \Ga$.
This is just a gauge fixing of the projective freedom which depends on the conformal structure $\gotC$ only, though $A$ depends on the representative $g$ chosen in $\gotC$; see also \cite{Dahidic}.
Locally one has
\[
\tilde\Ga^\al_{\be\mu}= \{g\}^\al_{\be\mu} + (g^{\al\ep} g_{\be\mu} -2\de^\al_{(\be}\de^\ep_{\mu)})A_\ep
\label{compatibilityCondition}
\]

Let us denote by $M$ the spacetime manifold and $\gotC$ and $\gotP$ two EPS-compatible structures on $M$. 
The triple $(M, \gotC, \gotP)$ is called an {\it EPS geometry}.
When the projective gauge has been fixed, the  triple $(M, \gotC, \tilde\Ga)$ is called a {\it Weyl geometry}.

Ehlers, Pirani and Schild attempted to further constrain EPS geometry by forcing the projective structure to be directly induced by the conformal structure. This can be done by requiring extra properties of worldlines. However, even Ehlers, Pirani, and Schild noticed that these extra assumptions appeared to be much less certain and less physically grounded than the other standard assumptions; see \cite{EPS}, \cite{EPS1}.

When there exists a conformal representative $\tilde g\in \gotC$ which also {\it represents} the projective structure (i.e.~the connection $\tilde\Ga=\{\tilde g\}$ is given by Levi-Civita connection of $\tilde g$) then the Weyl geometry is called {\it integrable}. 
In integrable Weyl geometries the $1$-form is exact, i.e.~$A\propto d \ln \Phi$, the potential $\Phi$ being directly related to the conformal factor which relates $g$ and $\tilde g$. 
In this case (and only in this case) the extra degrees of freedom in an EPS-compatible pair are encoded in just one scalar field $\Phi$.

In view of EPS framework it is therefore natural to use Palatini formalism to describe Gravity; metric and connection are considered {\it a priori} independent 
and while the metric $g$ is meant to represent light cones through its conformal structure, the connection $\tilde\Ga$ is meant to determine free falling of massive particles.
A priori, the connection is usually required to be torsionless (since torsion would not affect in any case the motion of test particles) but it is not restricted to be metric, even less to be determined by the metric $g$. 
Accordingly, the dynamics is expected to be conformally invariant (and of course {\it conformal transformations} are pointwise rescaling  of the metric, leaving the connection and the spacetime event fixed)  and to force the connection to be EPS-compatible with the conformal structure determined by field equations.

Let us stress explicitly the obvious: the metric $g$ contains more information than its conformal structure, since everything using a specific metric $g$ does rely on a gauge fixing of the conformal freedom. In particular distances and clock rates depend on such a conformal gauge fixing.

A number of dynamics with these properties are quite well known; see \cite{EPS2}.
There exists a class of couplings between gravity and matter (more precisely among the matter, the metric and possibly the connection $\tilde\Ga$) which force
the connection to be EPS-compatible with the conformal structure determined by $g$.

The relativistic theories where this happens are called {\it extended theories of gravitation} (ETG).
In such theories the connection is not even constrained to be metric, with all problems of holonomic nature which are well known (e.g.~the length of a ruler depends on its path!). EPS however provides a framework for the interpretation of gravitational theories and for discussing observability of the gravitational field in terms of motion of particles and light rays.

A subclass of ETG are {\it extended metric theories of gravitation} (EMTG) in which dynamics does not only constrain the connection to be EPS-compatible, but also to be metric, i.e.~$\tilde\Ga=\{\tilde g\}$ for some $\tilde g \in \gotC$. In these models there are no holonomic interpretation problems and there exists a single metric $\tilde g$ determining both light cones and free fall. These models correspond to the $1$-form $A$ determined to be exact (i.e.~$A\propto d \ln \Phi$) and they include all the so-called $f(\calR)$-theories in which the Lagrangian is assumed to be an analytic function of the scalar curvature $\calR= g^{\mu\nu} \tilde R_{\mu\nu}$ which depends both on the metric $g$ and the connection $\tilde\Ga$ (through its Ricci tensor).
In such a case the extra scalar is nothing but $f'(\calR)$. 

Standard General Relativity (GR) is obtained in the special case in which $\tilde\Ga=\{g\}$ which corresponds to a dynamics fixing $A=0$ (i.e.~$\Phi$ a constant normalized to be $1$).
We have to stress that choosing standard GR as the only possible model within a quite wide class of ETG
is not really reasonable even if it were eventually true. One should assume a more general attitude and then discuss in this wider framework if there are observational motivations to constrain the dynamics. Even if standard GR will eventually be confirmed by observation assuming {\it a priori} a definite model in Cosmology and Astrophysics is particularly problematic since most observations in these cases do depend on the model assumed and all sorts of hypotheses about the Physics to be described. It would be much better to rely on few assumptions and discuss possible dynamics on the bases of observations.

This attitude would be better even if  standard GR were perfectly describing data, at least to see if good fitting relies on physical reality or on (possibly unphysical) assumptions. Even better when we know that to save standard GR framework one is forced to add about 96\% of gravitational  sources (namely dark matter and dark energy) of which we have no direct evidence at fundamental level, which we know only through their gravitational effects, and at least part of which are quite questionable from the fundamental point of view; see  \cite{Faraoni}, \cite{Capozziello}, \cite{S2}, \cite{S3}, \cite{C1}, \cite{C2}, \cite{C4}. 
Moreover, Superstring effective theories do seem to imply modification of the Hilbert-Einstein Lagrangian of the same kind: see \cite{Odintzov}. 
The same result seem to be caused by LQG approach to Gravity; see \cite{Olmo}.

Since all Palatini $f(\calR)$-theories can be shown to be EMTG (see \cite{EPS1}, \cite{EPS2}), 
there is a single metric $\tilde g$ which is responsible for light cones and free fall.
One can try and write the whole model in terms of this metric from the very beginning; see \cite{Magnano}. 
As we shall briefly review below, this corresponds to a scalar-tensor theory which turns out to be in the particular form of a Brans-Dicke theory with parameter $\om=-3/2$
(and a suitable potential).
This Brans-Dicke theory is ruled out by experiments and this fact is often used to rule out Palatini $f(\calR)$-theories as well.
We shall argue that although the two models are mathematically equivalent as far as the variational principle is concerned, they are physically inequivalent due to extra assumptions about the free fall and possibly about observational protocols. 

In Palatini $f(\calR)$-theories one should consider whether observations are able to determine if we have defined distances by using $g$ or $\tilde g$. 
To this purpose one should consider that the conformal factor in $f(\calR)$ is determined by mass distribution.
Hence one should not expect any difference in vacuum (e.g.~in all solar systems test). 
Non-vacuum Einstein equations are used to model galaxies, clusters and Cosmology; notably in these cases standard GR fails to describe observations unless dark sources are introduced. In any event, in all this cases one should expect the conformal factor to be almost constant in space and in time.
Moreover, here in the solar system one knows that matter is almost irrelevant and the conformal factor is approximately $1$.
One could expect mismatches on distances at the scale of Planck which would be practically unobservable. 
But of course at cosmological scales we often perform observations at a distance of 10 billions of light years. At these scales effects may be observable and relevant.

Of course observational protocols are often disregarded in any variational approach to theoretical models. They need therefore to be made explicit.
First, we should notice that even in standard GR the issue of observability is quite poorly understood; see \cite{Rovelli}.
Since GR is generally covariant each observable quantity should be invariant with respect to diffeomorphisms.
Unfortunately, there are very few (if any) non-trivial generally covariant observables in GR. Even when observing standard geometric quantities such as volume and areas one has to be precise on what exactly is meant to be measured; see \cite{RovelliBook}.

Even a scalar, for example the scalar curvature $R$ of GR, is not generally invariant, since for a spacetime $\xi$ the Lie derivative  
 $\Lie_\xi R= \xi^\mu \del_\mu R$
does not need to be zero (unless one assumes $R$ to be constant, as it happens  for example in vacuum or when no matter other than electromagnetic field ---or any other traceless matter--- is present). Under this respect GR is quite different from other gauge theories in which gauge transformations are vertical and at least scalars are gauge invariant.

Thus one should wonder what is observed in Cosmology and Astrophysics? 
What we mean when we observe a supernova distant 3 billion light years, if that distance happens to be not generally invariant?  
According to Dirac framework for constrained theories such quantities should not be endowed with a physical meaning and nevertheless cosmologists and astrophysicists do routinely perform such measurements.
 
Rovelli proposed a framework in which observables are defined {\it against matter}; see \cite{Rovelli}.
For example, the scalar curvature $R$ at a spacetime point $x$ is not observable (in view of the hole argument) while the value of the scalar curvature $R$ at the intersections of two physical worldlines of two massive particles is in fact observable, as long as the diffeomorphisms drag both the scalar field $R$
{\it and the worldlines}. This seems to make sense out of what is measured in Astrophysics, though at the price of a careful review of
observational protocols.
A similar role of matter has been considered by EPS (the spacetime geometry is {\it built out} of matter worldlines).
Recently, we also considered a mechanism in which Weyl conformal invariance is broken by matter; see \cite{Fluidi1}, \cite{Marmo}, \cite{Corfu}.

In a private communication Pietro Menotti and Carlo Rovelli  pointed out to our attention that  standard tests of GR in Solar System can be performed
by using only angles (which are of course conformally invariant). 
We shall hereafter consider in detail the Mercury experiment using only angles, in order to trace precisely if and when the conformal invariance is broken.

The following is organized as follows: in Section 2 we shall review Palatini $f(\calR)$-theories and their equivalence with Brans-Dicke models.
In Section 3 we shall discuss the conformally invariant test of precession of perihelia.
In Section 4 we discuss in details how distance protocols break the conformal invariance.
 In Section 5 we consider a toy model in Cosmology to illustrate how conformal gauge has an influence on distances and observations.
The appendix is devoted to  discuss a simple mechanical example to show how similarities and differences between mathematical and physical equivalence may arise.

\

 \section{Palatini $f(\calR)$-theories and equivalence with Brans-Dicke theories}

Let us here review Palatini $f(\calR)$-theories.
Let $M$ be a spacetime manifold of dimension $4$ endowed with a metric $g$ and a (torsionless) connection $\tilde \Ga$ and let us consider a Lagrangian in the form
\[
L= \sqrt{g} f(\calR) + L_m(\psi, g)
\label{Lag}
\]
where   $\calR\equiv \calR(g, \tilde \Ga) := g^{\mu\nu} \tilde R_{\mu\nu}$, where $\tilde R_{\mu\nu}$ is the Ricci tensor of the independent connection $\tilde \Ga$,
where $f$ is a generic (analytic or {\it sufficiently regular}) function and $\psi$ is a collection of matter fields.

With this choice we are implicitly assuming that matter fields $\psi$ minimally couple to the metric $g$ which in turn encodes the `structural' electromagnetic properties of spacetime (photons and light cones). 
Of course, since the dynamics of the metric $g$ (which encodes causality and local Lorentz structure) and the dynamics of the connection $\tilde \Ga$ (which encodes free fall, i.e.~the interaction between gravity and point test particles) do {\it a posteriori} intertwine because of field equations, thence ---also {\it a posteriori}--- the dynamical behavior of matter will couple, along exact solutions, also to $\tilde \Ga$ and not only to $g$.  
To be more precise, in the Palatini framework the gravitational Lagrangian is of order zero in $g$ and of order one in $\tilde \Ga$, so that $g$ has no real dynamics  (its Euler-Lagrange field equations are zero order in $g$, i.e.~they express algebraic rather than differential conditions) while $\tilde \Ga$ has real dynamics (its Euler-Lagrange field equations have  order two in $\tilde \Ga$).            Because of coupling however, also $g$ obtains dynamics from the dynamics of $\tilde\Ga$, so that solving field equations for   $\tilde\Ga$  (if possible) will provide non-trivial dynamics to the field equations of $g$ as well.

It would probably be better to be more liberal and allow matter couplings to the connection (see \cite{EPS2}, \cite{Olmo}, \cite{MGaCou}).
Let us here notice that what follows can be  in fact extended to a more general framework; one can in fact work out classes of matter Lagrangians depending on the connection $\tilde\Ga$ in which field equations still imply the EPS-compatibility condition (\ref{compatibilityCondition}); see \cite{EPS2}, \cite{S1}, \cite{S2}.
However, also in view of simplicity, the  matter Lagrangian $L_m$ is here assumed to depend only on matter and metric. 
Field equations of (\ref{Lag}) are then
\[
\begin{cases}
f'(\calR) \tilde R_{(\mu\nu)} -\Frac[1/2] f(\calR) g_{\mu\nu}=\ka T_{\mu\nu}
\qquad \(T_{\mu\nu}=\Frac[1/\sqrt{g}]\Frac[\de L_m/\de g^{\mu\nu}]\)\\
\tilde \na_\al \left(\sqrt{g} f'(\calR) g^{\be\mu}\right)=0\\
\end{cases}
\label{FEQ}
\]
where $f'(\calR)$ denotes the derivative of the function $f(\calR)$ with respect to its argument $\calR$.
We do not write the matter field equations which will be considered as matter equations of state.
The constant $\ka= 8\pi G/c^4$ is the coupling constant between matter and Gravity.

Under these simplifying assumptions the second field equation can be solved explicitly by introducing a ($\tilde\Ga$-dependent) conformal transformation $\tilde g_{\mu\nu}= f'(\calR)\cdot g_{\mu\nu}$.
As a consequence the connection is eventually given as 
$\tilde \Gamma^\al_{\be\mu} = \{\tilde g\}^\al_{\be\mu}$, 
i.e.~the connection $\tilde\Ga$ is forced to be the Levi-Civita connection of the conformal metric $\tilde g$.
Thus in these theories the connection is {\it a posteriori} metric and the geometry of spacetime is described by an integrable Weyl geometry; see \cite{EPS2}.
The trace of the first field equation (with respect to $g^{\mu\nu}$) is so important in the analysis of these models that it has been called the {\it master equation};
see \cite{Universality}.
It reads as
\[
f'(\calR) \calR -2 f(\calR) =\ka T
\label{msterEQ}\]
where $T= g^{\mu\nu}T_{\mu\nu}$ is the trace of the energy-mementum tensor $T_{\mu\nu}$.
For a generic (sufficiently regular) function $f$, the master equation establishes, as we said, an {\it algebraic} (i.e.~not differential) relation for $g$ which can be solved for $\calR=r(T)$.
The first field equation becomes then
\[
\tilde G_{\mu\nu}= \tilde R_{\mu\nu} - \Frac[1/2] \tilde R \tilde g_{\mu\nu}
=\ka\left(\Frac[1/\vp(T)] \left(T_{\mu\nu}- \Frac[1/4 ] Tg_{\mu\nu}\right)
- \Frac[1/4]  \hat r(T)  g_{\mu\nu} \right)=:\ka \tilde T_{\mu\nu}
\label{Palatini f(R)}
\]
where we set $\vp(T)=f'(r(T))$.
If the trace  $T$ is not constant but is a genuine spacetime function, then the conformal factor $\vp(T)$ is not constant.
 Accordingly we see that a Palatini $f(\calR)$-theory with matter behaves like standard GR for the conformal metric $\tilde g $ with a  modified source stress tensor.
Naively speaking, one can reasonably hope that the modifications dictated by the choice of the function $f$ can be chosen to fit observational data; see \cite{Faraoni}, \cite{C1},
\cite{S3}, \cite{Capozziello}, \cite{C4}.

In a sense, whenever $T\not=0$ in presence of standard visible matter $\psi$, an energy momentum stress tensor $T_{\mu\nu}$ would produce by gravitational interaction with $\tilde \Ga$ (i.e.~with the conformal metric $\tilde g= \vp(T)\cdot g$) a kind of {\it effective} energy-momentum stress tensor $\tilde T_{\mu\nu}$
in which standard matter $\psi$ is seen to exist together with {\it dark (virtual)} matter generated by the gauging of the rulers imposed by the $T$-dependent conformal transformations on $g$. In a sense, the {\it dark side} of Einstein equations can be mimicked by suitably choosing $f$ and $L_m$, as a curvature effect induced by $T= g^{\mu\nu} T_{\mu\nu}\not=0$; see \cite{C1}, \cite{C2}, \cite{C4}. 

In vacuum or for purely electromagnetic matter obeying Maxwell equations, Palatini $f(\calR)$-theories are generically equivalent to Einstein models with cosmological constant  and the possible values of the cosmological constant form a discrete set which depends on the analytic function $f$.
This is known as the {\it universality theorem} for Einstein equations (see \cite{Universality}).

One can consider field equations (\ref{Palatini f(R)}) and the master equation (\ref{msterEQ}) back in the original metric $g$ obtaining
\[
\begin{cases}
\vp R_{\mu\nu}= \na_{\mu\nu} \vp +\Frac[1/2]\Dal\vp g_{\mu\nu}-\Frac[3/2\vp] \na_\mu\vp\na_\nu \vp + \Frac[\ka/4] \vp  r(T) g_{\mu\nu}
		+    \ka \left(T_{\mu\nu} -\Frac[1/4]Tg_{\mu\nu}\right)\\
\vp R= 3 \Dal\vp -\Frac[3/2\vp] \na_\al \vp \na^\al\vp + \ka T + 2 f\\
\end{cases}
\label{f1}\]

Within the framework for  $f(\calR)$-theory one can generically invert for the conformal factor
$\vp= f'(\calR)$ 
to obtain $\calR= \si(\vp)$ 
and define a potential function
\[
U(\vp)= -\vp \si(\vp)+ f(\si(\vp))
\qquad (\Rightarrow\>U'(\vp)= -\si' \vp - \si + f' \si'= -\si)
\label{Potential}
\]
 
Equations (\ref{f1})  can be thence recognized as field equations of a {\it Brans-Dicke} theory for a metric $g_{\mu\nu}$ with dynamics described by a Lagrangian in the following form
\[
L_{BD}= \sqrt{g}\left[ \vp R - \Frac[\om/\vp] \na_\mu \vp \na^\mu \vp + U(\vp)\right] + L_m(g, \psi)
\label{BDAction}\]
where we set $\om=-\Frac[3/2]$.
Thus we can summarize the situation by saying that any Palatini $f(\calR)$-theory is equivalent to a Brans-Dicke theory (with $\om=-\Frac[3/2]$ and a suitable potential).
Let us remark that Brans-Dicke theories (without potential) are considered in testing standard GR; see \cite{Weinberg}. In fact standard GR corresponds to the limit $\om\arr \infty$ and classical tests within Solar System rule out small values of $\om$ (among which $\om = -3/2$ as for $f(\calR)$-theories).
Thus we have two  theories which are mathematically equivalent (one can map the action, field equations and solutions of one into the other by a `conformal transformation')
one of which (Brans-Dicke) is ruled out by observations.
Can we conclude that the other theory (Palatini $f(\calR)$-theory) is ruled out as well?

In Appendix A we discuss the issue in a simple mechanical example. 
To answer let us remark that, first of all, what is ruled out by observations is Brans-Dicke {\it without a potential}. 
One should then discuss whether the potential has some influence on observation. Let us also remark that the potential (\ref{Potential}) is singular exactly on standard GR where
the conformal factor $\vp= f'(\calR)\equiv 1$ cannot be solved for $\calR$.
For the sake of argument let us assume that the potential does not affect observations and Brans-Dicke model (\ref{BDAction}) {\it is} ruled out.

Secondly, there is a number of features in a field theory which are assumed independently of the action principle as independent assumptions.
One is the interpretation of physical quantities which cannot be directly derived by the action principle. 
In Brans-Dicke theory the free fall is dictated by the metric $g$, in Palatini $f(\calR)$-theories with EPS interpetation it is dictated by $\tilde g$.
In the two {\it ``mathematically equivalent''} models bodies fall along different worldlines. For example Mercury will go along different orbits so that, as we shall see below, the perihelia precession test is failed by Brans-Dicke theories though this will not apply to  Palatini $f(\calR)$-theories. We shall be back on this test below to discuss it in further detail.

Moreover, also an independent assumption has to be done about which metric should be used to define distances in spacetime. In Palatini $f(\calR)$-theories one has two natural conformal metrics (and in fact a whole conformal class). Each representative of the conformal class defines a different notion of distance and it is quite hard to see which metric is selected by the {\it usual} observational protocols. This issue has been noticed by Ehlers Pirani and Schild as well; they stop to discuss how firmly we know that {\it gravitational time} (which is what we call the proper time) is identical to {\it atomic time} (which is what we use) and they concluded that the issue cannot be easily addressed; see also \cite{Perlick}.  

Finally, let us stress that we are using {\it conformal transformation} with two different meanings. In EPS framework a conformal transformation consists in changing the metric, leaving the connection (as well as the spacetime point) unchanged. In view of EPS analysis these transformations are expected to be gauge transformations since one cannot observe representatives of the conformal structure.
When discussing the equivalence with Brans-Dicke theories (see \cite{Magnano}) we made a {\it ``conformal transformation''} to go back to the original metric $g$. 
However, at that point we already eliminated the connection $\tilde \Ga=\{\tilde g\}$ so that by acting on the metric $\tilde g$ we {\it also} act on the connection.
This is not a gauge transformation as the one found in EPS and in fact this affects the model.

\section{Conformal view on Mercury precession}

Let us now consider the test for precession of perihelia of Mercury for $f(\calR)$-theories.
First we need a model for the gravitational field around the Sun; that is well approximated by a static, spherically symmetric vacuum solution.
In view of the universality theorem (see \cite{Universality}), we {\it know} that the metric $\tilde g$ must be some sort of Schwarzschild-(A)dS solution and the conformal factor is a constant related to the cosmological constant. We know by experience that cosmological constant effects within the Solar System are hard to be detected so that we expect the solution to be well approximated by Schwarzschild solution. 
In view of the EPS interpretation Mercury, unlike in Brans-Dicke theories, goes along the geodesics of the Scharwschild $\tilde g$.
Thence one expects $f(\calR)$-theories to be almost identical to standard GR and quite different from Brans-Dicke, at least in this situation.
It is then natural to expect that a whole family of $f(\calR)$-theories will be able to pass the precession test as well as standard GR.
Unfortunately, by the same argument it is natural to expect to be difficult to test Palatini $f(\calR)$-theories against standard GR.

It is instead interesting to go through the Mercury test in the context of Palatini $f(\calR)$-theories, i.e.~in view of EPS interpretation,  tracing the influence of conformal transformations. 
In particular, let us go through it at first by relying only on conformally invariant quantities. 
This will provide insights about the meaning of the conformal factor.


Let us suppose we are suddenly teleported to a star system (e.g.~by using {\it the Machine} of the novel {\it Contact}).
The star system is made of a quite compact star (named Sun2, {\it S2} for short) and a single planet (called Mercury2, {\it M2} for short) orbiting around S2.
The star system does not appear to be in a galaxy but is floats in outer space away form influences of other bodies.
There is also an abandoned space station orbiting the star farer away from M2 in (what will turn out to be) a perfect circular and perfectly coplanar orbit, always facing S2.

The observation space station kindly left after by some alien civilization is of course called Earth2 ({\it E2} for short) and it is set up to perform two measurements:
one can measure the angle $\be$ between a fix star on the ecliptic plane and S2 (or equivalently to measure which fix star is at right angle with S2 so staying away from possible deflections of lightrays) and to measure the angle $\al$ between M2 and S2.

\begin{figure}[htbp] 
   \centering
   \includegraphics[width=2.5in]{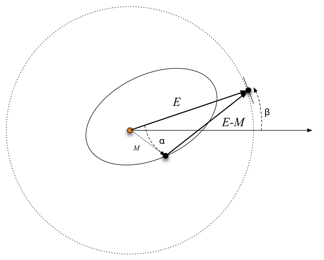} 
   \caption{Orbit of M2 and E2 around S2}
\end{figure}

Then one can obtain a dataset made of pairs of angles $(\be, \al)$, or equivalently the pairs $(\be, \cos(\al))$.
Being the readings angles they are invariant in the conformal class: regardless the representative which is chosen for the conformal class $[g]$, being it $g$ or $\tilde g= \Om^2 \cdot g$, the angles do not change.

Let us assume, for the sake of argument, that a single orbit is well described by Kepler laws one can predict a graph for the function $\cos(\al)= \chi(\be)$.
Let us neglect at first the time needed for light to propagate.
If one denotes by $\vec E$ the vector from S2 to E2 and by $\vec M$ the vector from S2 to M2, then the angle $\al$ is obtained as
\[
\cos(\al)= \Frac[\vec E\cdot (\vec E-\vec M)/ \Vert \vec E\Vert \> \Vert \vec E-\vec M\Vert]\equiv \chi(\vp)
\]

The vector $\vec E$ is directly observed as $\vec E= r_E(\cos\be \> \vec i + \sin\be\> \vec j)$, and the vector $\vec M$ is given by $\vec M= x\> \vec i + y\>\vec j$ where we set
\[
\begin{cases}
X:= a \cos\vp  + d\\
Y:= b \sin \vp\\
\end{cases}
\qquad\qquad
\begin{cases}
x= \cos\te X + \sin\te Y\\
y= -\sin\te X + \cos\te Y\\
\end{cases}
\]
Here $\vp$ is the angle between M2 and the center of its elliptical orbit with respect to a fixed direction. 
Because of first Kepler law $a$ and $b$ are the semiaxes of the orbital ellipse of M2 and $d$ is the focal distance from the center. 
If we introduce the eccentricy $\ep$ then one has $d= \ep a$ and $b=a\sqrt{1-\ep^2} $. 
Let us also denote by $\te$ the rotation angle of the orbital ellipse with respect to a fixed direction.

Because of the second Kepler law, angular momentum of M2 is conserved. This can be used to eliminate the time dependence of the orbit
\[
k_M= b(a+d\cos(\vp))\dot \vp
\qquad\then
t=\Frac[b/k_M](a\vp + d\sin\vp)
\label{time}\]
and parametrize the graph $\cos(\al)=\chi(\be)$ by $\vp$.

Let us set $\ell r_E= a$; being $\ell$ the ratio of two distances it is conformally invariant as an angle.
By the third Kepler law, the orbital constant of M2 is equal to the orbital constant of  E2.
Then
\[
 \Frac[4\pi^2 a^2b^2 /\ell^3a^3k_M^2] =   \Frac[4\pi^2/\om_E^2r_E^3]
\qquad\then
\om_E^2 = a\Frac[k_M^2  \ell^3/b^2 r_E^3]= \Frac[k_M^2  \ell^2/(1-\ep^2) r_E^4]
\]
 
Let us first notice that $\be=\om_E t=  \sqrt{\ell^3}(\vp + \ep\sin\vp)$ is independent of $k_M$.
The angle $\be$ can be used as a time coordinate.
Now both $\vec E$ and $\vec M$  (and then the function $\chi(\vp)$) are expressed in terms of the parameter $\vp$. 

Second, it is easy to show that the function $\chi$ turns out to depend on the conformally invariant parameters $(\ell, \te,\ep)$
but it is independent of the scale parameters $(r_E, k_M)$. In particular one has
\[
\chi(\vp; \ep, \te, \ell)= \Frac[\ell-(A+B)/\sqrt{\ell^2+\(1+\ep\cos(\vp)\)^2-2\ell \(A+ B\)}]
\]
where we set
\[
\begin{cases}
\xi= \ell^{-3/2}\(-\vp\( 1-\ep^2+\ep^2\cos(\te)\) -\ep\sin(\vp)\cos(\te)+\ep \sqrt{1-\ep^2}\cos(\vp)\sin(\te)\) \\
A= \cos(\xi)\(\cos(\te)(\ep+\cos(\vp))+\sin(\te)\sqrt{1-\ep^2}\sin(\vp)\)\\
B= \sin(\xi)\(\sin(\te)\( \ep+\cos(\vp)\)-\cos(\te)\sqrt{1-\ep^2}\sin(\vp)\)\\
\end{cases}\]

Then we obtained a parametric representation of the graph $\cos(\al)$ as a function of $\be$ given by
\[
\begin{cases}
\be= \sqrt{\ell^3}(\vp + \ep\sin\vp)\\
\cos(\al)= \chi(\vp; \ep, \te, \ell)\\
\end{cases}
\]
Then one can encode the Kepler prediction into this curve that can be compared directly with the observed dataset.

\begin{figure}[htbp] 
   \centering
   \includegraphics[width=2.5in]{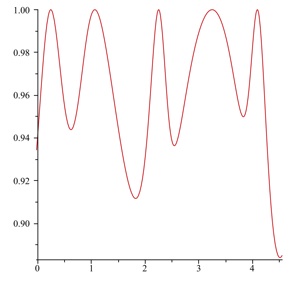} 
   \caption{Prediction of $\cos(\al)=\chi(\be)$ using Kepler laws}
\end{figure}

First of all by fitting the dataset against the curve one can obtain an estimate of the parameters and in principle a lot of redundancy
to test the hypothesis that M2 motion is well described by  Kepler approximation.
Let us stress that the test is completely conformally invariant. We did not use rulers or clocks at any event. 
We are using the orbital motion of E2 as a clock but in fact we are measuring only angles.

Let us suppose for the sake of argument that we find a good fitting on few orbits with Kepler laws and the parameters
\[
\ell\simeq 0.380
\qquad
\ep\simeq 0.205
\]
as well as value $\te\simeq \te_0 $ for the orientation of M2 orbit.
By a pure coincidence the eccentricity is about the eccentricity of the orbit of Mercury and $\ell$ is about the ratio of the Earth's and Mercury's orbital radii  in our Solar System.

Now that we know that Kepler is a good approximation and we have values for orbital parameters of M2,
we can repeat the measurement over and over and find that $\ep$ and $\ell$ are constant over time within the measurement errors while $\te$ appear to be increasing linearly along a line. 
This shows that we are able to check over time (or $\be$) the evolution of the perihelia of M2. 
For the sake of argument let us suppose we find that $\te$ increases of about an angle $\De \te=2.086\cdot 10^{-4}$rad every 100 revolutions of E2, i.e.~in the range $0\le \be\le 200\pi$.

\subsection{Speed of light}

To be more precise one should take into account that light travels at a finite speed $c$.
At a given time $t_0$ E2 is at position $\vec E$, M2 is at position $\vec M$.
M2 emits a spherical light front which propagates at speed $c$ and reaches E2 at time $t_1$ at position $\vec E_1$.
At that point from E2 the planet M2 is seen at the position $\vec M$ where it emitted the light.
This time lapse produces an aberration of the angle $\al$ that slightly modify the predicted curve 
$\cos(\al)= \hat\chi(\be)$.

One can compute the difference $\hat\chi(\be)-\chi(\be)$.
First, one can see the correction to be quite small. 
Second, one can see that once again the correction as a function of $\be$ is conformally invariant (while it would depend on scale if considered as a function of time, which, unfortunately, we have no way of measure).

Once again being the correction conformally invariant, one cannot use these corrections to determine the scale.

\begin{figure}[htbp] 
   \centering
   \includegraphics[width=2in]{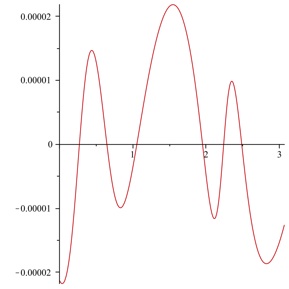} 
   \caption{Corrections due to light propagation as a function of $\be$}
\end{figure}

\subsection{Conformal invariance}

It is pretty clear that the star system we are  studying is very similar to (a simplified version of) the Solar System.
Unfortunately, for some reason we have no clock to measure the revolution period of E2 or a ruler to measure $r_E$ (or $a=\ell r_E$). 
Without a  clock we have a conformally invariant result which is unable to fix a global scale.
We defined a model with no time, in which time is replaced by the angle $\be$ giving a beautiful example of how naturally Leibniz {\it relational time} appears in Astronomy. 
Let us stress that this is exactly what astronomers measure when using AU units. 
The conformal invariance is then broken (or gauge fixed) only when one states the equivalence $1AU= (1.4959870700\cdot 10^{11} \pm 3)\>m$.
Let us remark that this gauge fixing is done by measuring  the speed of light, which is done on Earth and now.

In some sense we could say that relativistic theories are naturally with no scale which is in fact encoded, partially as a convention, when we decide to use Newtonian time.
In fact one can see that the curve showing $\cos(\al)$ as a function of $t$, by taking into account (\ref{time}), does in fact depend on the scale through 
the  parameters $a$ and $k_M$ appearing in (\ref{time}). Hence, if we had a clock, by fitting these curves we would obtain also the scale and all non-conformally invariant orbital parameters.

Since in Astronomy the mass of the Sun is {\it defined} by the third Kepler law (and it is in fact a  non-conformally invariant orbital parameter) to be $GM= 4\pi^2a^3 T^{-2}$
where $a$ and $T$ are the orbital semiaxe and period, respectively, for any planet (e.g.~Mercury).
As a result, with no clock, i.e.~in a Weyl geometry, we are not able to distinguish between our Solar System and one larger, slower, with a more massive star, provided that distances, periods, and masses are rescaled by the same factor.
This is true until we get and independent way of measuring the inertial mass of the Sun (that we cannot).


One could think that relativistic effects could break the conformal invariance, but this is not the case.
Let us remark that standard GR (as well as Palatini $f(\calR)$-theories since the prediction just depends on the fact that free fall is dictated by Schwarzschild metric) 
predicts
\[
\De \te= \Frac[6\pi / 1-\ep^2 ]\Frac[{GM_{\hbox{\tiny Sun}}}/a] 
\quad\hbox{ (rad per Mercury revolution)}
\]
i.e.~$\De\te \simeq 43.03"$ per 100 revolutions of the Earth.
In view of the definition of the mass of the Sun, the quantity   $GM_{\hbox{\tiny Sun}}/a$ is conformally invariant, as of course is the factor $6\pi / (1-\ep^2)$.

Within Brans-Dicke (with no potential) theory, Mercury moves along different orbit and the prediction would be
different by a factor 
\[
\De \te= \Frac[6\pi / 1-\ep^2 ]\Frac[ GM_{\hbox{\tiny Sun}}/a] \left(\Frac[3\om+ 4/3\om +6]\right)
\quad\hbox{ (rad per Mercury revolution)}
\]
which is sensibly different from the prediction of both standard GR and for Palatini $f(\calR)$-theories (by a factor $-1/3$ if one takes $\om=-3/2$).
Thus to summarize both standard GR and Palatini $f(\calR)$-theories (unlike Brans-Dicke theories) predict a shift of perihelia of Mercury of 
\[
\De \te= 43.03"
\quad\hbox{ (rad per 100 revolutions of the Earth)}
\]

Such a prediction is conformally invariant (while of course it would not be so if stated in {\it rad per century} which would depend on a clock fixing).
The conformal invariance tells us that until we fix a clock (or a ruler) we are not able to distinguish on observational stance between our Solar System 
and a star system which happens to be bigger, slower and with a star which is more massive than the Sun by a single factor which rescaled distances time and masses.
If we misjudged distances and times for some reasons this implies extra gravitational mass of the star.

\section{Units of  distances}

Until the conformal factor is constant one can change the units of distances, time lapses, masses (as well as the universal constants), 
to compensate exactly the effect. Imagine now that on the space station we have an alien clock which tics but we cannot gauge it against the second.
The only thing one can do is to invent a new unit for measure, the {\it tic}($\tau$) which corresponds to the pace of the alien clock.  

We easily check that the revolution period of E2 is $T_E\simeq 3.156\cdot 10^7\>\tau$ (i.e.~1 year {\it if} a tic were a second).  
We can also echo a light ray  on S2 and  get the signal back after about $1000\>\tau$. 
Let us define a length unit called a {\it step} ($\sigma$) to be the distance $1/299792458$ of the distance travelled by light in $1\>\tau$.
Accordingly, the  speed of light is $c= 299792458 \>\si/\tau$. (If $c$ were the same we know and a tic were a second then a step would be 1 meter.)

Since clocks cannot travel through the Machine we are fundamentally unable to fix the scale of our star system.
We can only assume that $1\>\tau= \la \> s$ (and possibly we can guess that $\la \sim 1$ though we cannot easily test the error).
Because of our definition of step and $c$ we also have that $1\>\si= \la \> m$. For simplicity let us assume that $\la\lsim 1$.

As discussed above, we can perform classical and relativistic Astronomy in a conformally invariant way.
Such a procedure will confirm Kepler's laws and will allow to observe precession of perihelia of planets.
Unfortunately, one cannot fix the scale in this way. An exact copy of the Solar System will be indistinguishable from a smaller and faster star system where length, revolution periods and the mass of the central star are rescaled by the same factor.

Since we observe distances in Astronomy, since we declare that the Earth orbit is between $147\cdot 10^6 km$ and $152\cdot 10^6 km$ that means that we usually consider observable more than what we can observe by conformally invariant protocols.
When we say that the distance between the Earth and the Sun is $147\cdot 10^6 km$ we are stating a certain ratio between this distance and the standard ruler. Since we are basically measuring the time needed by light to travel from the Sun to the Earth, we are basically taking as a foundation for our definition of distance the direct measurement of two ways light velocity done in the last few centuries on the Earth (for example resorting to the celebrated Fizeau-Foucault experiment).

Of course the distances are also related to the metric we decide to use to represent  the conformal structure, simply because we fix it to be the metric which allows us to compute distances in spacetime as Minkowski metric in Special Relativity.
Whenever one discusses about metrics in a generally covariant theory one has also an awkward relation with coordinates (which are conventional though they allow to write down the metric in components). Usually the convention in GR is that the coordinates have the dimension of a  length, while the metric coefficients are adimensional; we shall get stuck to this notation here.

When in $f(\calR)$ theories we have two conformal metrics in the game then we have a conformal transformation transforming (actively) one metric
into the other. At the same time we have a change of coordinates transforming (passively) the local representation of both metrics. Probably is a good idea to write down explicitly the relations involved before discussing the definition of distances.

Let us consider two sets of coordinates; one $x^\mu$ adapted to meters and seconds, the other $x'^\mu=\la x^\mu$ adapted to steps  and tics. 
The alien sample ruler starts at coordinate $x'=0$ and ends at coordinate $x'=1\si= \la^{-1} m$.
A terrestrial ruler would start at $x=0$ and would end at $x= 1m= \la \si $.
Let us also fix the conformal factor to be $\Phi^2=\la^{2}$.

We can summarize the whole situation in the following diagram:
\[
\begindc{\commdiag}[1]
\obj(10,20)[gx]{$g=-\(1-\Frac[2GM/r]\)dt^2 + \Frac[1/{1-\Frac[2GM/r]}] dr^2+r^2 d\Om^2$}
\obj(260, 20)[gx']{$g=\la^{-2}\left[-\(1-\Frac[2GM'/r']\)dt'^2 + \Frac[1/{1-\Frac[2GM'/r']}] dr'^2 +r'^2 d\Om^2\right]$}
\obj(10,90)[tgx]{$\tilde g= \la^{2} \left[-\(1-\Frac[2GM/r]\)dt^2 + \Frac[1/{1-\Frac[2GM/r]}]dr^2+r^2 d\Om^2\right]$}
\obj(260, 90)[tgx']{$\tilde g=-\(1-\Frac[2GM'/r']\)dt'^2 + \Frac[1/{1-\Frac[2GM'/r']}] dr'^2+r'^2 d\Om^2$}
\mor{gx}{gx'}{\tiny passive}
\mor{tgx}{tgx'}{\tiny passive}
\mor{gx}{tgx}{\tiny active}
\mor{gx'}{tgx'}{\tiny active}
\enddc
\]

We can easily figure out that we use the conformal gauge fixing 
\[
g=-(1-\Frac[2GM/r])dt^2 + \Frac[1/{1-\Frac[2GM/r]}] dr^2+r^2 d\Om^2
\] 
while the alien civilization uses
\[
\tilde g=-(1-\Frac[2GM'/r'])dt'^2 + \Frac[1/{1-\Frac[2GM'/r']}] dr'^2+r'^2 d\Om^2
\]
In other words it does not matter which metric corresponds to the operational definition of distances, one can always define his units for length
so that one see it as a Schwarzschild metric.

Thus one could argue that what we are considering is not very physical and that everything amounts to a redefinition of units.
However, this is not the case if we consider regions of spacetime in which the conformal factor is not approximately constant.
In that case the effect of the conformal factor cannot be exactly cancelled by a unit redefinition; see \cite{OlmoNostro}.

\section{Conclusions and Perspectives}

We discussed here how the interpretation of gravity as in EPS framework implies a Palatini or metric-affine formalism which is conformally invariant
(with a special kind of conformal transformations which rescales the metric leaving connection and coordinates unchanged).

A family of natural candidates are Palatini $f(\calR)$-theories which are almost indistinguishable from standard GR. 
In vacuum (or with electromagnetic field only) any Palatini $f(\calR)$-theory is equivalent to a standard Einstein theory with a cosmological constant
(which experimentally we can presume to be small).
The effects are confined in matter (where by the way we have evidences of exotic behavior of Gravity and where our knowledge 
is limited). 

To be precise we treated EPS interpretation in general and assumptions about free fall in particular as independent of the variational principle.
As a matter of fact free falls of test particles should be derived and related to characteristics of matter field equations. However, one can regard to these relations as a constraint on gravitational matter couplings.

Although Palatini $f(\calR)$-theories are mathematically equivalent to Brans-Dicke theories the two models are definitely not the same theory on physical grounds. 
The two theories are connected by a conformal transformation (of the kind which affects the metric and the connection) however in
Palatini $f(\calR)$-theories free fall is dictated by $\tilde g$, while in Brans-Dicke theories it is dictated by $g$.
As a consequence, Palatini $f(\calR)$-theories pass Mercury test, while Brans-Dicke theories do not.

Besides these effects which are related to dark sources, we also possibly have  a different, somehow independent, family of  effects.  
These new effects are linked to our operational definition of distances, time lapses and the physical meaning of the conformal factor.
These effects are dumped when the conformal factor is constant, but they  appear when its changes become relevant.
When we consider two regions with different values of the conformal factor, one could think of canceling the effects in each region by a change of units
at the price of using different units in different regions. 
For example in Cosmology the conformal factor is naturally related to the density of matter which is decreasing during expansion.
It is {\it as if} units for distances, times lapses and masses (as well as the universal constants) were time dependent, 
though their changes are not arbitrary but they are constrained by their common relation with the conformal factor and the choice of the function $f$.

Since the observational protocol are quite obscure in their origin it is particularly hard to see directly whether these units mismatched are really there but one should assume them, make prediction, test them and falsify the existence of these mismatches. Let us just here notice that accelerations are not conformally invariant, thus it is at least conceivable 
that the acceleration of the Universe expansion may depend on the conformal factor.

Finally, let us remark that although we showed that Palatini $f(\calR)$-theories pass Mercury test and argue they should pass the classical Solar System tests,
nevertheless this does not mean that there is no effect able to falsify them.
One should consider each and single observational protocol and test it in view of the richer Weyl geometry and see it as a constraint on the dynamics related to the specific choice of $f$,
possibly eventually confirming the standard dynamics from $f=\calR$.
For example we have quite strong evidence of the structure of emission lines from very far away atoms (as well as constraints from nucleogenesis).
For example we know that one cannot change the fine structure constant without  it becoming evident in the emission spectrum, which gives quite a good though indirect control on the value of the constant in early Universe. The control on the gravitational constant $G$ is much more relaxed (being on the order of $10\%$ which leave us with a huge family of possible $f$ to be explored). In any event in Palatini $f(\calR)$-theories one always has two theoretical definitions, one using $g$ and one using $\tilde g$, for each quantity and one should always investigate which is the theoretical quantity which better fits the experimental data, leading to a better understanding of the meaning of quantities measured in Astrophysics and Cosmology.

\section[A]{Appendix: Mathematical Equivalences in Physics}

Let us discuss a simple mechanical example to highlight how mathematical equivalence cannot be extended to a full physical equivalence.
Here the point is that even in view of a mathematical equivalence which in principle allows to map one description into the other, 
observational protocols  may break the equivalence. This may happen expecially when observation are difficult and one is unable to measure  everything but just some quantities can be measured, as it happens in Astrophysics and Cosmology.

\begin{figure}[htbp] 
   \centering
   \includegraphics[width=4cm]{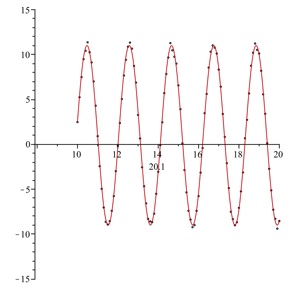} 
   \qquad
   \includegraphics[width=4cm]{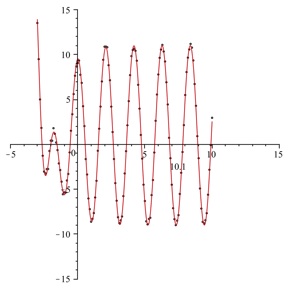}    
   \caption{$x(t)$ in the time range a)  $t\in[10,20]$. \qquad  
   		b)  $t\in [-3,10]$}
\end{figure}

Imagine we are given a material point constrained on a straight line free to move under unknown forces to be studied. 
We know that one can measure the position  $x$  and the the momentum $\pi$.
We can easily plot $x$ and $\pi$ in their time evolution. 

One question which is easily asked is whether the system can be described as a Hamiltonian system. For it, one should identify 
two quantities $q$ and $p$, give a Hamiltonian function $H(q, p)$ and relate the evolution of the system and the solutions of the Hamilton equations
\[
\begin{cases}
\dot q= \Frac[\del H/\del p]\\
\dot p=- \Frac[\del H/\del q]\\
\end{cases}
\]

By observing the motion of the point in the range (a) $t\in [10,20]$ one could make an educated guess for a harmonic oscillator.
By a closer look one could fit the data very well by a function
\[
x=  A\cos(\om (t+t_0)) + \la
\]
with $ A=10, \om= 3, \la=1, t_0=0$.
The system is thence described by the Hamiltonian
\[
\bar H= \Frac[1/2] p^2 + \Frac[\om^2/2] q^2 -\la\om^2 q
\]
Accordingly, one can guess the force to be a harmonic force plus a constant force.

After that one obtains a bigger dataset $(b)$ which includes the time range $t\in[-3,10]$.
The new dataset clearly disagrees with previous guesses.  
It seems that something awkward happened at time $t\sim 0$. One could either suppose some extra temporary force acted and it is then switched off, or
look for as single time dependent force explaining both datasets.

In fact a very good fit can be obtained by the function
\[
x=\al(t)(A\cos(\om(t-t_0))+\la)
\]
for a parameter function $\al(t)= 1-e^{-(t+2)}$ and again with $A=10, \om= 3, \la=1, t_0=0$.  

This is a solution of Hamilton equations of the following Hamiltonian
\[
H=  \Frac[\al^2/2] p^2 + \Frac[\om^2/2\al^2] q^2 -\la\Frac[\om^2/\al] q + \Frac[\dot \al/\al] pq
\]
Of course this model is a big step forward since it perfectly describes the whole dataset.
On the other hand terms in $pq$ has not a direct mechanical  interpretation.

Then one can try to look for a canonical transformation to simplify the model.
In particular one can define new (time-dependent family of) canonical coordinates
\[
\begin{cases}
Q= \Frac[q/\al(t)] -\la\\
P= \al(t) p\\
\end{cases}
\]
and check that the new Hamiltonian is simply 
\[
K= \Frac[1/2] P^2 +\Frac[\om^2/2] Q^2
\]
Obviously, the canonical transformation establishes a very well founded mathematical equivalence between the two Hamiltonian systems described by $H$ and $K$ (which is in fact {\it one} Hamiltonian system with two different local representations).

To what extent though, the system is a pure harmonic oscillator on the physical stance?
To answer the question maybe it is worth considering some remarks. 
First, the dynamics of $K$ is much simpler than the dynamics of $H$.
Second, in the $K$ framework there is an observable $x= \al(t) (Q +\la)$ which fits the dataset as perfectly as $q$ does in the $H$ framework.
Third, in the $H$ framework the observable used for fitting is simply $x= q$.

Accordingly, we have two mathematically equivalent frameworks, one in which the dynamics is particularly simple, the other in which what we observe
(i.e.~our observational protocols) is particularly simple.  Because of the particular situation either we find a way of observing directly the quantity $Q$ (which would make the $K$ framework superior under all viewpoints) or we have to resign to have two frameworks each simpler under a different viewpoint.

Let us also finally remark that the transformation between the two frameworks, namely $q= \al(t)(Q+\la)$, can be directly related to a (time-dependent) mismatch of the protocol of measuring the position. It is as if, besides changing the origin $\la$ of the position reference frame, we did change the unit of distances by a (time-dependent) factor $\al(t)$ (imagine for example we are using ultrasounds to define distances and the speed of sound in changing with time
as it would happen if the experiment were held in a space lab during a decompression.
As another example, let us  imagine we are using a conformal metric to define distances and the conformal factor is depending on the matter density 
which is changing in a ever expanding universe).
The mismatch does not emerge in datasets covering intervals in which the function $\al$ can be considered constant (as it happens in the dataset $a$)
while it becomes manifest once the dependence of $\al$ on time can be appreciated.

Though we do not need to recall that in Cosmology the protocols for defining and measuring distances are quite obscure from a fundamental perspective and depend on many aspects of the underlying model and many physical assumptions, we are not here claiming that a similar mechanism can explain cosmological observations.
However, we believe that one should understand in detail how and why this is not the case.

This rather trivial example in any event shows how one should not use mathematical equivalence to dismiss a framework without carefully reviewing the observational protocols and verifying their compatibility with the equivalence transformations, especially in Cosmology where we know from the very beginning that most of the time we are measuring quantities that are not gauge covariant.

\section*{Acknowledgements}

We wish to thank P.Menotti, C.Rovelli and O.Bertolami for discussions and comments.
We also acknowledge the contribution of INFN (Iniziativa Specifica NA12) the local research project 
{\it Metodi Geometrici in Fisica Matematica e Applicazioni} (2011) of Dipartimento di Matematica of University of Torino (Italy).
This paper is also supported by INdAM-GNFM.

\end{document}